# Atomic-scale identification of the active sites of nanocatalysts

Yao Yang[1*], Jihan Zhou[1*], Zipeng Zhao[2*], Geng Sun[3*], Saman Moniri[1], Colin Ophus[4], Yongsoo Yang[1], Ziyang Wei[5], Yakun Yuan[1], Cheng Zhu[6], Qiang Sun[7], Qingying Jia[7], Hendrik Heinz[6], Jim Ciston[4], Peter Ercius[4], Philippe Sautet[3,5], Yu Huang[2], Jianwei Miao[1†]

[1]*Department of Physics & Astronomy and California NanoSystems Institute, University of California, Los Angeles, CA 90095, USA.* [2]*Department of Materials Science and Engineering, University of California, Los Angeles, CA 90095, USA.* [3]*Department of Chemical and Biomolecular Engineering, University of California, Los Angeles, Los Angeles, CA 90095, USA.* [4]*National Center for Electron Microscopy, Molecular Foundry, Lawrence Berkeley National Laboratory, Berkeley, CA 94720, USA.* [5]*Department of Chemistry and Biochemistry, University of California, Los Angeles, Los Angeles, CA 90095, USA.* [6]*Department of Chemical and Biological Engineering, University of Colorado at Boulder, Boulder, CO, USA.* [7]*Department of Chemistry and Chemical Biology, Northeastern University, Boston, MA, USA*

[*]These authors contributed equally to this work.

[†]Corresponding author. Email: miao@physics.ucla.edu (J.M.)

**Alloy nanocatalysts have found broad applications ranging from fuel cells to catalytic converters and hydrogenation reactions. Despite extensive studies, identifying the active sites of nanocatalysts remains a major challenge due to the heterogeneity of the local atomic environment. Here, we advance atomic electron tomography to determine the 3D local atomic structure, surface morphology and chemical composition of PtNi and Mo-doped PtNi nanocatalysts. Using machine learning trained by density functional theory calculations, we identify the catalytic active sites for the oxygen reduction reaction from experimental 3D**

**atomic coordinates, which are corroborated by electrochemical measurements. By quantifying the structure-activity relationship, we discover a local environment descriptor to explain and predict the catalytic active sites at the atomic level. The ability to determine the 3D atomic structure and chemical species coupled with machine learning is expected to expand our fundamental understanding of a wide range of nanocatalysts.**

Identification of active sites is crucial for understanding the properties of heterogeneous catalysts and for rational design of improved catalytic activities (*1, 2*). Despite significant progress from various experimental and computational methods (*3-8*), localization of the active sites of nanocatalysts remains largely elusive, particularly in multicomponent nanoparticles. This limitation is mainly due to an incomplete understanding of the three-dimensional (3D) atomic arrangement of the different constituents and the structural reconstruction driven by catalytic reactions (*7, 9-11*). For the electrochemical oxygen reduction reaction (ORR) – a limiting half-reaction in fuel cells (*12*), the single-crystal $Pt_3Ni$ (111) facet has been demonstrated to exhibit a very high ORR activity (*13*). Yet, a fundamental understanding of the catalytic activity of Pt-alloy nanocatalysts is far more challenging due to the heterogeneity of the catalytic active sites. Various experiments have indicated that the structural morphology and chemical composition such as facets, surface concaveness, strain and ligand effects can all play a role in the ORR activity (*14-20*). However, correlating the 3D atomic structure and chemical composition with the active sites of the Pt-alloy nanocatalysts remains a difficult task. On the computational side, density functional theory (DFT) is a powerful method to predict the ORR activity by incorporating crystal defects into perfect lattices and using relaxed atomic configurations at the minimum energy states (*3, 21*). However, real nanocatalysts neither have perfect crystal lattices nor are always in the minimum energy states in their active configuration. Here, we applied atomic electron tomography (AET) (*22-26*) to determine the 3D atomic coordinates of 17 PtNi and Mo-doped PtNi (Mo-PtNi) nanocatalysts before and after activation. We identified the facets, surface concaveness, structural and chemical order/disorder, coordination numbers (CN), and bond lengths with unprecedented 3D atomic detail. The experimental 3D atomic coordinates were coupled with DFT-trained machine learning (ML) to identify the active sites of the nanocatalysts, which were further validated by electrochemical measurements.

PtNi and Mo-PtNi nanocatalysts with varying Ni concentration were synthesized on carbon black / nanotubes using an efficient one-pot approach (Fig. S1, Tables S1 and S2) (*20, 27, 28*). To investigate the 3D atomic structure after activation, a fraction of the nanocatalysts were activated with 30 cycles of cyclic voltammetry. The ORR specific activities of PtNi BA, PtNi AA, Mo-PtNi BA and Mo-PtNi AA were measured to be 2.9, 4.8, 3.3, 9.3 mA/cm$^2$ at 0.9 V$_{RHE}$, respectively (Fig. S2 and Table S3), where AA and BA represent before and after the activation. The increase of the ORR activity with Mo dopants and the high specific activity of Mo-PtNi AA are consistent with previous reports (*20, 27, 29*). AET experiments were performed on 17 nanocatalysts using an annular dark-field scanning transmission electron microscope (Tables S1, S2 and Figs. S3-6). After pre-processing, 3D reconstruction, atom tracing and refinement (Fig. S7) (*23, 24, 28, 30, 31*), the 3D atomic coordinates and chemical species of the 17 nanocatalysts were determined (Fig. 1), where the pixel size was calibrated by an extended x-ray absorption fine structure measurement (Table S4) (*28*). Due to a very small fraction of Mo dopants (1.4% for BA and 0.4% for AA) (*11*), AET is presently not sensitive enough to distinguish them from the Ni or Pt atoms (*23, 24*), but this limitation will not impact the conclusions of this study.

Figure 1A shows the 3D surface morphology and the chemical composition of four representative nanocatalysts (PtNi BA, PtNi AA, Mo-PtNi BA and Mo-PtNi AA). Elemental segregation was observed on the surface and in the interior of the nanoparticles (Fig. 1A and B). The surface layer mainly consists of Pt atoms, forming (100), (110), (111) and high-index facets (Fig. 1C). From the experimental 3D coordinates, we quantitatively characterized the surface concaveness, structural and chemical order/disorder, CN, and surface bonds at the single-atom level (Fig. 1C-H) (*28*). The 3D atomic structure and the chemical composition of the other 13 nanocatalysts are shown in Figs. S8 and S9. Although the majority of the nanoparticles exhibit an octahedral morphology, we observed surface concaveness, structural and chemical disorder to varying degrees in all four types of the nanocatalysts. We found that the Mo dopants and the activation increase the surface concaveness and the structural disorder (Fig. S10). From the experimental 3D atomic coordinates, we calculated the mean and the standard deviation of the surface Pt-Pt bond length to be 2.73±0.12 Å, 2.77±0.16 Å, 2.71±0.19 Å and 2.75±0.19 Å for PtNi BA, PtNi AA, Mo-PtNi BA and Mo-PtNi AA, respectively (Fig. 2A and B). The slight increase of the mean surface Pt-Pt bond length after activation is attributed to the loss (leaching) of the surface and subsurface (that is, the layer below the surface) Ni during the activation.

Next, we analyzed the CN of the surface Pt atoms, where CN = 9 has the highest fraction for all four types of nanocatalysts (Fig. 2C and D). We also examined the nearest-neighbor Pt and Ni atoms of each surface Pt atom, termed $NN^{Pt}$ and $NN^{Ni}$, respectively. Figure 2C and D shows the distribution of $NN^{Pt}$ and $NN^{Ni}$ for the four types of nanocatalysts. The activation shifted the distribution of $NN^{Pt}$ of the nanocatalysts to the right, and the distribution of $NN^{Ni}$ to the left. This observation reveals that activation leached more Ni than Pt atoms from the surface and subsurface. Figure 2E shows the average surface and subsurface Ni composition for the four types of nanocatalysts. The activation reduced the surface and subsurface Ni composition by 15.4% and 23.1% for PtNi and by 22.6% and 8.2% for Mo-PtNi, respectively, showing that Mo-PtNi preserved more subsurface Ni atoms during activation. We also observed a correlation between the subsurface Ni composition and the surface Pt-Pt bond length (Fig. 2F). With the increase of the subsurface Ni composition, the average surface Pt-Pt bond length decreases for all the four types of nanocatalysts, indicating that subsurface Ni increases the compressive strain of the surface Pt sites.

From the experimentally determined 3D atomic coordinates, we used a DFT-trained ML method to identify the ORR active sites of the nanocatalysts. The ORR takes place mainly through a four-step electroreduction $O_2 + 4\ (H^+ + e^-) \rightarrow 2H_2O$, in which OH is an intermediate (*23, 24*). Extensive DFT studies reveal that the ORR activity follows the Sabatier principle and that optimal catalysts have the OH binding energy ($BE_{OH}$) about 0.1-0.15 eV weaker than that of bulk Pt(111) (*4*). As it is computationally impractical to perform DFT calculations for all the surface sites of the nanocatalysts, we utilized DFT-trained ML to predict the OH binding energy for the experimentally measured surface Pt sites (*28*). We first constructed 207 3D PtNi atomic models each surrounding a surface Pt site with a different local environment. After calculating the $BE_{OH}$ for the 207 Pt sites by DFT, we randomly chose 134 sites to train the ML method and then used it to identify the $BE_{OH}$ of the 73 test Pt sites. A quantitative comparison between the DFT calculated and ML identified $BE_{OH}$ is shown in Fig. S11, indicating that ML accurately predicted the $BE_{OH}$ with a root mean square error (RMSE) of 0.05 and 0.07 eV per site for the 134 training and 73 test Pt sites, respectively.

After training and validating the ML method, we applied it to evaluate the ORR activity for the experimentally measured surface Pt sites of the PtNi AA and Mo-PtNi AA nanocatalysts. We focused on the activity of the AA nanoparticles because the 3D morphology and structure of the BA nanoparticles were likely modified by the ORR test performed in 0.1 M $HClO_4$ (*7, 9*). The ML method was used to estimate the $BE_{OH}$ of the 26,246 surface Pt sites for the 7 PtNi AA and 4 Mo-Pt/Ni AA nanocatalysts. By referring $BE_{OH}$ to the OH binding energy of Pt(111), we derived the ORR activity for all the surface Pt sites (*4, 28, 32, 33*). The average catalytic activities of the PtNi AA and Mo-PtNi AA nanocatalysts agree with the electrochemical measurements (Fig. 3A), showing the robustness of using DFT-trained ML to identify the ORR activity from the experimental 3D atomic coordinates. Figure 3B, C and Fig. S12 show the ORR activity map of the surface Pt sites of the PtNi AA and Mo-PtNi AA nanocatalysts. A striking feature is the difference of the ORR activity of the surface Pt sites by several orders of magnitude. While the majority of the surface Pt sites have a very low catalytic activity, there are a very small fraction of highly active sites (yellow atoms in Fig. 3B and C). Figure 3D-I shows six representative highly active sites from the PtNi AA and Mo-PtNi AA nanocatalysts, each of which exhibits a distinct 3D local environment such as different CN, nearby Ni atoms and surface morphology. This observation indicates that quantitative characterization of the 3D local atomic environment is critical to the understanding of the active sites of nanocatalysts.

The combination of the experimentally determined 3D atomic coordinates and the ML results enabled us to perform a comprehensive analysis of the structure-activity relationship and to identify a descriptor for the surface Pt sites (*28*), which we termed the local environment descriptor (LED). LED is dimensionless like the CN and defined as,

$$LED = NN^{Pt} \cdot e^{-a_1 \cdot \varepsilon} + a_2 \cdot \overline{CN}^{Ni}, \quad (2)$$

where $\varepsilon = \frac{\bar{d}_{Pt} - d_0}{d_0}$ is the local strain with $\bar{d}_{Pt}$ the average Pt-Pt bond length around a surface Pt site and $d_0$ the Pt-Pt bond length (2.75 Å) for Pt nanoparticles, $\overline{CN}^{Ni} = \sum_i \frac{CN_i^{Ni}}{CN_{max}}$ is the generalized CN of the considered Pt with Ni atoms (*3, 34*), $CN_i^{Ni}$ is the CN of the $i^{th}$ Ni atom, $CN_{max} = 12$ is the CN in bulk Pt, $a_1$ and $a_2$ are two fitting constants related to the strain and ligand effects, respectively. Our quantitative analysis showed that $NN^{Pt}$, $\bar{d}_{Pt}$ and $\overline{CN}^{Ni}$ are more relevant to the activity of the nanocatalysts than other factors such as surface concaveness, structural and chemical order/disorder (*28*). By fitting LED to the $BE_{OH}$ of all the surface Pt sites relative to the OH binding energy of Pt(111), we obtained a volcano-type activity plot with the peak at LED = 9.7 (Fig. 3J). The RMSE of the fitting is 0.104 eV with $a_1 = 1.985$ and $a_2 = 1.075$. Figure 3K, L and Fig. S13 show the activity map of the PtNi AA and Mo-PtNi AA nanocatalysts based on LED, which is in agreement with that identified by the ML method (Fig. 3B, C and Fig. S12).

To gain insight into LED, we considered a simple case. For Pt(111) without strain, we calculated LED = 9 from Eq. (1), which is on the left-side of the peak of the volcano plot (Fig. 3J). If a nearest-neighbor surface Pt atom is substituted by a Ni atom, the first term of Eq. (1) decreases by 1, but the second term increases by a number smaller than 1, making LED smaller than 9. If the substitutional Ni atom is in the subsurface, the CN of the Ni is 12 and the second term increases by a number larger than 1 as $a_2 > 1$, making LED larger than 9. Furthermore, compressive strain, induced by the decrease of the average Pt-Pt bond length, also increases LED. Consequently, with LED < 9.7, both subsurface Ni and compressive strain increase the ORR activity. When LED reaches the peak of the volcano plot (LED = 9.7 in Fig. 3J), further increasing the subsurface Ni and compressive strain reduces the ORR activity. This analysis demonstrates the importance of balancing the strain and ligand effects to optimize the ORR activity of PtNi nanocatalysts.

Our experimental results reveal the differences of the average surface Pt-Pt bond length and the standard deviation between the Mo-PtNi AA (2.75±0.19 Å) and PtNi AA (2.77±0.16 Å) nanocatalysts (Fig. 2A and B), showing that the former has a larger compressive strain than the latter. Additionally, Mo-PtNi AA preserves more subsurface Ni atoms than PtNi AA (Fig. 2E).

According to Eq. (1), both factors increase the LED of Mo-PtNi over PtNi. As the measured and the ML-identified ORR activity of Mo-PtNi are higher than those of PtNi (Fig. 3A), our observations indicate that the average activity of the Mo-PtNi AA and PtNi AA nanocatalysts is situated on the left-side of the peak of the volcano plot (Fig. 3J), which can explain previous experimental results that a larger concentration of Co or Ni in Pt-alloy nanocatalysts increases the ORR activity (*35, 36*). Although we focused on PtNi nanoparticles in this study, LED is in principle applicable to other bimetallic nanocatalysts.

In conclusion, we determined the 3D local atomic structure and chemical species of PtNi and Mo-PtNi nanocatalysts before and after activation, and measured the facets, surface concaveness, structural and chemical order/disorder, CN, and bond lengths with high precision. From the experimentally measured 3D atomic coordinates, we used a DFT-trained ML method to identify the active sites of 11 nanocatalysts, which were validated by electrochemical measurements. We observed that the ORR activity of the surface Pt sites of the nanocatalysts varies by several orders of magnitude. By performing a quantitative analysis of the structure-activity relationship, we discovered a descriptor (dubbed LED) to understand the ORR activity of the nanocatalysts based on the surface, subsurface atomic structure and chemical composition. We found that the nearest-neighbor surface Pt atoms, the average Pt-Pt bond length and the generalized CN for Ni neighbors are most relevant to the ORR activity and contribute to LED. The optimal reactivity is achieved with the right balance between the ligand and strain effects, with subsurface Ni ligands behaving differently from surface ones. We anticipate that this general method can be used to measure the 3D local atomic positions, chemical species and surface composition of a wide range of nanocatalysts for various (electro)chemical reactions and to understand their structure-activity relationships at the single-atom level.

**ACKNOWLEDGMENTS**: This work was primarily supported by the US Department of Energy (DOE), Office of Science, Basic Energy Sciences (BES), Division of Materials Sciences and Engineering under award DE-SC0010378. It was also partially supported by STROBE: A National Science Foundation Science & Technology Center under grant number DMR 1548924 and the NSF DMREF under award number DMR-1437263. G.S. and P.S. acknowledge the support by DOE-BES grant DE-SC0019152. AET experiments were performed with TEAM I at the Molecular Foundry, which is supported by the Office of Science, Office of Basic Energy Sciences of the US DOE under contract number DE-AC02-05CH11231. The XAS experiment was conducted on beamline 8-ID (ISS) of the National Synchrotron Light Source II, which is supported by the Office of Science, Office of Basic Energy Sciences of the US DOE under contract number DE-SC0012704.

**FIGURES**

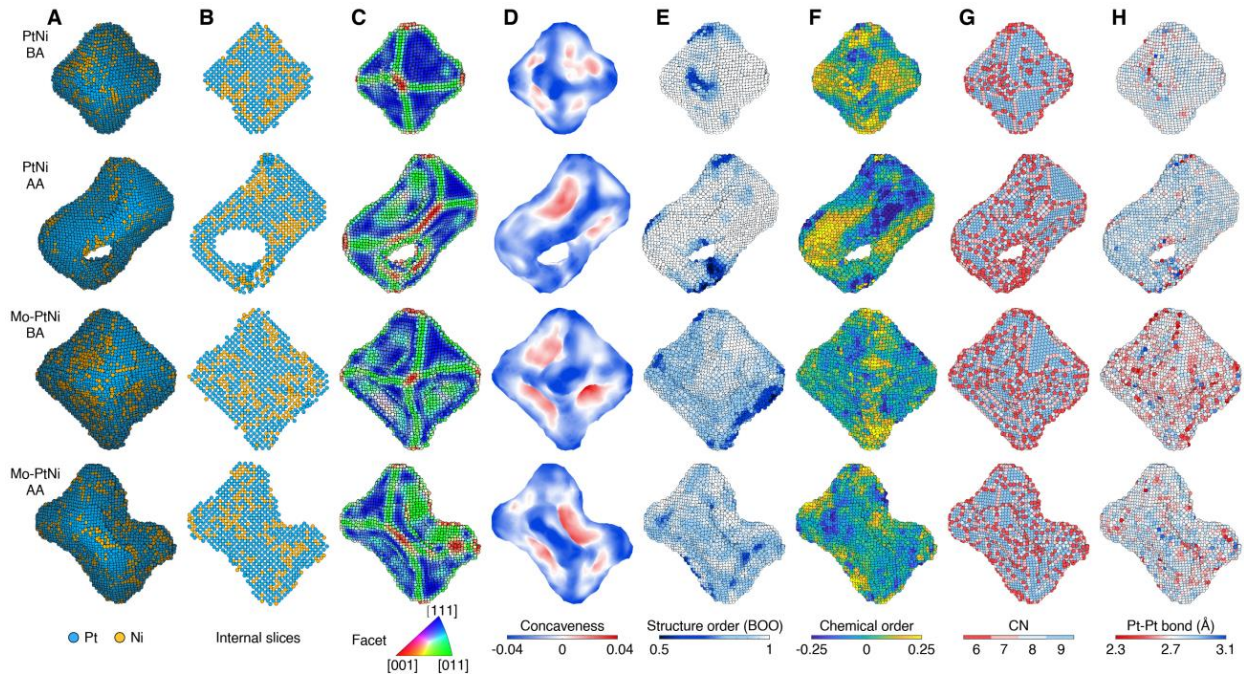

**Fig. 1. 3D atomic structure and chemical composition of four representative nanocatalysts determined by AET.** From the experimental atomic coordinates, the 3D surface morphology and chemical composition (**A**), elemental segregation in the interior (**B**), facets (**C**), surface concaveness (**D**), structural order/disorder (**E**), chemical order/disorder (**F**), CN (**G**), and surface bonds (**H**) of the nanocatalysts were identified at the single-atom level. For the structural order, BOO = 1 corresponds to a perfect fcc lattice. For the chemical order, positive and a negative numbers represent segregation and alloying, respectively.

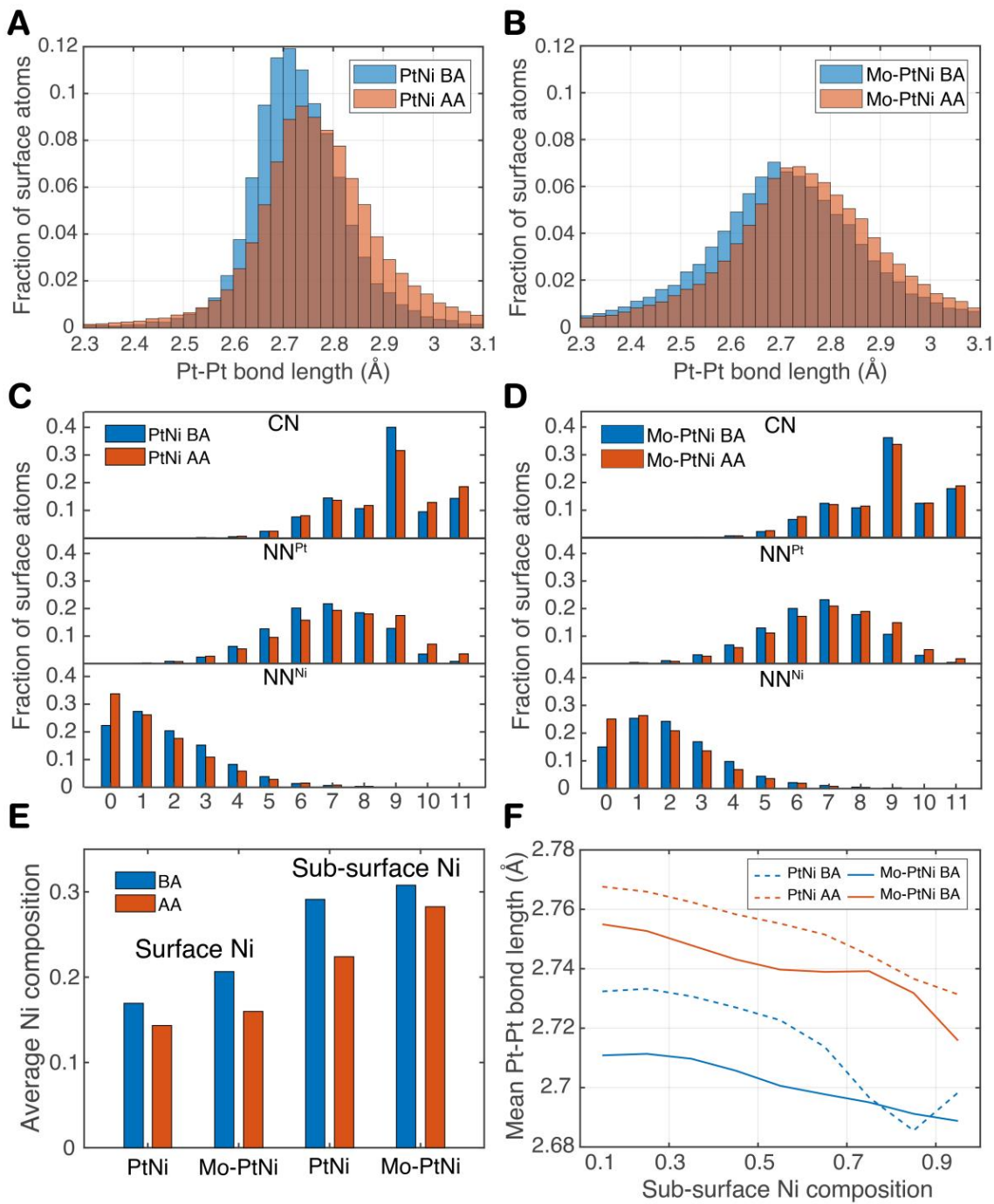

**Fig. 2. Quantitative characterization of the four types of nanocatalysts.** Distribution of the surface Pt-Pt bond length for PtNi BA and AA (**A**), and Mo-PtNi BA and AA (**B**). Histogram of the CN, NN$^{Pt}$ and NN$^{Ni}$ of the surface Pt sites for PtNi BA and AA (**C**), and Mo-PtNi BA and AA (**D**). (**E**) Histogram of average surface and subsurface Ni composition. (**F**) Correlation between the subsurface Ni composition and the surface Pt-Pt bond length.

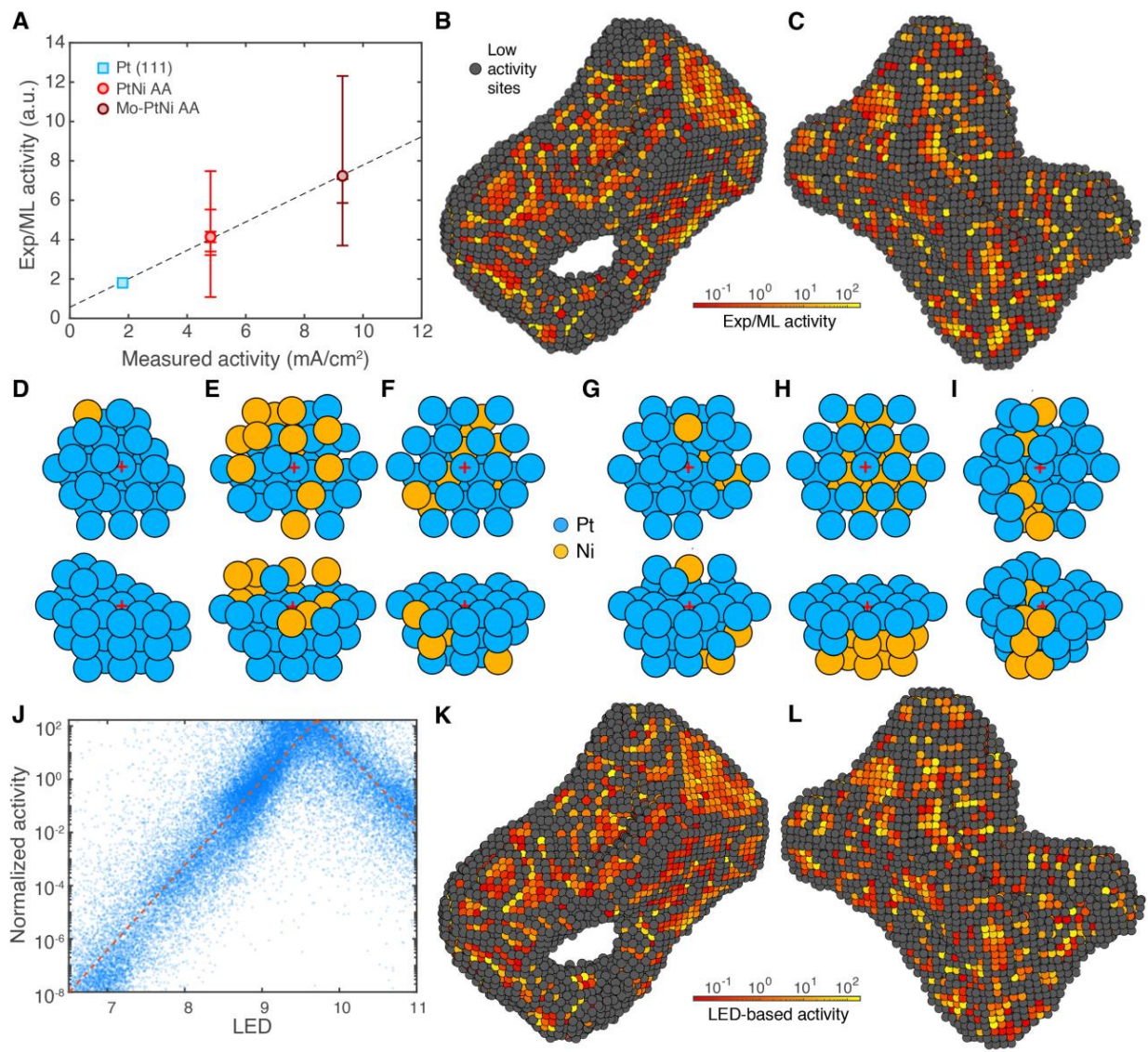

**Fig. 3. Identification of the active sites of nanocatalysts.** (**A**) Comparison between the electrochemically measured ORR activity and the ML-identified activity from the experimental 3D atomic coordinates of PtNi AA and Mo-PtNi AA, where each bar represents a nanocatalyst and the circle the average activity. The activity of Pt(111) was obtained from DFT as a reference point. (**B** and **C**) The activity distribution of the surface Pt sites of the PtNi AA and Mo-PtNi AA nanocatalysts shown in the second and fourth row of Fig. 1, respectively, where low activity sites are defined with the ORR activity smaller than 3% of that of Pt(111). Representative highly active sites (red crosses) from the PtNi AA (**D-F**) and Mo-PtNi AA (**G-I**) nanocatalysts, showing different 3D local environments. (**J**) Volcano-type activity plot (red dashed line) obtained by fitting LED with the ML-identified activity of all the surface Pt sites (blue dots), where the peak is at LED = 9.7. (**K** and **L**) The activity distribution of the two PtNi AA and Mo-PtNi AA nanocatalysts based on LED, respectively, which is in agreement with that identified by ML (**B** and **C**).

# Supplementary Materials

**Materials and Methods**

Chemicals and Materials

Platinum(II) acetylacetonate [Pt(acac)$_2$], nickel(II) acetate tetrahydrate [Ni(ac)$_2$·4H$_2$O], benzyl acid (BA) were purchased from Sigma Aldrich. Molybdenum hexacarbonyl (Mo(CO)$_6$), carbon nanotube (CNT) was purchased from Alfa Aesar. N, N-Dimethylformamide (DMF), acetone, isopropanol were purchased from Fisher Scientific. Ethanol was purchased from Decon Labs, Inc. Vulcan XC-72 carbon black (particle size ~50 nm) was from Cabot Corporation. Water used was Ultrapure Millipore (18.2 MΩ·cm).

Sample preparation

*Synthesis of Mo-PtNi/C*. Vulcan XC-72 carbon black is pretreated in Argon (80% in volume) and Hydrogen (20% in volume) mixture at 400 C for 4 hours. 80 mg pretreated Vulcan XC-72 carbon black was dispersed in 60 mL N,N-dimethylformamide (DMF) under ultrasonication for 30 minutes in a 325 mL pressure bottle. Then 64 mg platinum(II) acetylacetonate [Pt(acac)$_2$], 32 mg nickel(II) acetate tetrahydrate [Ni(ac)$_2$·4H$_2$O], and 520 mg benzoic acid were dissolved in 10 mL DMF and were also added into the 325 mL pressure bottle with carbon black dispersion. After ultrasonication for 5 mins, the pressure bottle with well mixed solution was directly put into 140 ºC oil bath and then slowly heated to 160 ºC within 2 hrs. The pressure bottle was kept at 160 ºC for 12 hrs. After 12 hours, 16 mg Pt(acac)$_2$, 8 mg Ni(ac)$_2$·4H$_2$O, 3.2 mg molybdenum(0) hexacarbonyl [Mo(CO)$_6$] were added into the pressure bottle. Then the pressure bottle was kept in 160 ºC oil bath for 48 hrs. After reaction finished, the catalysts were collected by centrifugation, then dispersed and washed with isopropanol and acetone mixture. Then the catalysts were dried in vacuum at room temperature and ready for characterization and electrochemistry test.

*Synthesis of PtNi/C*. The preparation of procedure is same as Mo-PtNi/C noted above except without adding Mo(CO)$_6$.

*Synthesis of Mo-PtNi/CNT*. The preparation procedure is same as Mo-PtNi/C noted above except replacing treated Vulcan XC-72 with CNT.

*Synthesis of PtNi/CNT*. The preparation procedure is same as Mo-PtNi/CNT except without adding Mo(CO)$_6$.

Electrochemical measurements

A typical catalyst ink was prepared by mixing 2.8 mg of catalyst powder (Mo-PtNi/C, PtNi/C) with 2 mL of ethanol solution containing 16 μL of Nafion (5 wt%) with 5 min ultrasonication time. Then, 10 μL of catalyst ink was dropped onto a 5 mm diameter glassy-carbon electrode (Pine Research Instrumentation). Estimation of Pt loading is based on overall Pt ratio within catalyst determined by ICP-AES. The ink was dried under an infrared lamp; then the electrode was ready for electrochemical test.

A three-electrode cell was used to carry out the electrochemical measurements. The working electrode was a catalyst coated glassy carbon electrode. A Ag/AgCl electrode was used as the reference electrode. A Pt wire was used as the counter electrode. Cyclic voltammetry (CV) activation was conducted in a N$_2$ saturated 0.1 M HClO$_4$ solution between 0.05 to 1.1 V vs.

reversible hydrogen electrode (RHE) at a sweep rate of 100 mV/s for 30 cycles. Oxygen reduction reaction (ORR) measurements were conducted in an $O_2$ saturated 0.1 M $HClO_4$ solution at a sweep rate of 20 mV/s. iR compensation and background subtraction are applied for ORR test curves following the protocol noted in literature (*37*). For the ORR measurement without activation, the prepared working electrode was directly subjected to ORR test in the oxygen saturated electrolyte without being activated at nitrogen saturated electrolyte via CV scan.

Data acquisition

The PtNi and Mo-PtNi nanoparticles were deposited on to 5-nm-thick silicon nitride membranes annealed at 520 °C for 9 minutes in vacuum. A set of tomographic tilt series were acquired from several nanoparticles using the TEAM I microscope. Images were collected at 300 kV in ADF-STEM mode (Table S1 and S2). To minimize sample drift, three to five images per angle were measured with 3 μs dwell time. To ensure that no structural changes were observed during the data acquisition, for each nanoparticle, we took the same projection images at zero degree before, during, and after the acquisition of each tilt series. 9 PtNi and 8 Mo-PtNi nanoparticles were measured in this work. The total electron dose of each tilt series for all the nanoparticles were estimated to be between $7.4 \times 10^5$ $e^-/Å^2$ and $8.5 \times 10^5$ $e^-/Å^2$ (Table S1 and S2).

Image pre-processing

The image pre-processing consists of the following three steps.

i) *Image registration and drift correction*. We acquired three to five ADF-STEM images at every angle of each tilt series. The images at each angle were registered by cross-correlation. Using first image as a reference, we scanned a cropped region of the subsequent images with a sub-pixel step size and found the drift for every image. After drift correction, we averaged all the images at each angle.

ii) *Image denoising*. The experimental ADF-STEM images have mixed Poisson and Gaussian noise. A generalized denoising algorithm, termed block-matching and 3D filtering (BM3D), was used to denoise each averaged image (*38*). The robustness of BM3D on the AET data have been proven in our previous studies (*23, 24, 30, 31, 39*).

iii) *Image alignment and background subtraction*. The denoised images of each tilt series were aligned by the center of mass and common line method as described elsewhere (*23, 31*). After alignment, a 2D mask was calculated by the Otsu threshold in MATLAB for each image to match the shape of the nanoparticle. The background was estimated by the discrete Laplacian function in MATLAB. After background subtraction, all the images of each tilt series were re-aligned by the center of mass and common line to further reduce the error.

3D image reconstruction and refinement

After image pre-processing, each tilt series was reconstructed by an iterative algorithm, termed REal Space Iterative REconstruction (RESIRE) (*31, 39*). From the experimental images, RESIRE minimized the $L_2$-norm error metric using gradient descent. The $j^{th}$ iteration of the algorithm consists of the following steps. RESIRE computed a set of images from the 3D object of the $(j-1)^{th}$ iteration. The difference between the computed and corresponding experimental images was calculated, from which an error metric was defined to monitor the convergence of the algorithm. The difference was back projected to real space, yielding the gradient of the 3D reconstruction. The 3D reconstruction of the $j^{th}$ iteration was updated by combining the gradient with the reconstruction of the $(j-1)^{th}$ iteration, where positivity was enforced as constrains. As a general algorithm, RESIRE is not sensitive to the initial input. After 200 iterations, a preliminary 3D reconstruction was computed. Based on the preliminary 3D reconstruction, angular refinement and spatial alignment were performed and background subtraction was re-evaluated. After these refinement procedures, a final 3D reconstruction was obtained by running 200 iterations of RESIRE.

Determination of 3D atomic coordinates and chemical species

To increase the precision of atom tracing, we up-sampled each 3D reconstruction by a factor of three using spline interpolation, from which all the local maxima were identified. Starting from the highest intensity, we fit each local maximum of a 9×9×9 voxel volume (1.4×1.4×1.4 Å$^3$) by a 3D polynomial method to locate its center position (*23, 24, 31*). Each fitted local maximum was considered as a potential atom only when its distance from the existing potential atoms is larger than 2 Å. After going through all the local maxima, we obtained a list of potential atoms. For every potential atom, the integrated intensity of the 9×9×9 voxel volume around the center position was calculated. A K-mean clustering method was used to classify the non-atoms, Pt and Ni atoms (*31*). Due to a small fraction (<1.5%) of Mo atoms in the Mo-PtNi nanoparticles, AET is currently not sensitive enough to distinguish them from Pt or Ni atoms (*27*). After excluding the non-atoms and manually checking all the atoms, we obtained an initial 3D atomic model for each 3D reconstruction.

Due to the missing wedge problem (*22*), we used local re-classification to reduce the effect of the intensity variation in different regions of each 3D reconstruction (*23*). At each atomic position, we cropped a 7-Å radius sphere and calculated the mean integrated intensity for the Pt or Ni atom inside the sphere. We then re-classified each atom in the sphere based on the difference between its integrated intensity and the mean value of the Pt or Ni atom. The procedure was repeated until there was no further change.

The electron energy loss spectroscopy maps of the nanoparticles show that there are individual Ni atoms distributed around each nanoparticle (Fig. S1). To evaluate the effect of the surrounding Ni atoms on the 3D reconstruction, we simulated a PtNi atomic model in an environment with individual Ni atoms, which used the experimentally determined 3D atomic distribution for the Pt and Ni atom. After calculating projection images at different tilt angles from the model, we performed image pre-processing, conducted the 3D reconstruction, traced the atoms, classified the atomic species and obtained a new 3D atomic model (Fig. S7). We observed that there is a layer of ghost atoms surrounding the 3D atomic model, which is due to the surrounding Ni atoms around each nanoparticle (Fig. S7). Based on this result, a layer of ghost atoms was removed from the experimental 3D atomic model of each nanoparticle.

X-ray absorption spectroscopy (XAS) data collection and analysis.

XAS experiments were conducted on the dry powders of the four nanocatalysts studied in this work at the beamline ISS 8-ID in National Synchrotron Light Source II (NSLS) (Brookhaven National Laboratory, NY. Full range Pt L$_3$-edge and Ni K-edge spectra were collected on the same electrode in transmission mode at the Pt L$_3$-edge, and/or Ni K-edge, with a Pt or Ni reference foil positioned between I2 and I3 as a reference. Typical experimental procedures were utilized with details provided in our previous work (*40*). The data were processed and fitted using the Ifeffit−based Athena (*41*) and Artemis (*42*) programs. Scans were calibrated, aligned and normalized with background removed using the IFEFFIT suite (*43*). The χ(R) were modeled using single scattering paths calculated by FEFF6 (*44*).

The Pt-Pt bond lengths of the four nanocatalysts were obtained by the extended x-ray absorption fine structure (EXAFS) fitting (Table S4). The average first-shell Pt-Pt bond lengths were determined by fitting the EXAFS spectra of dry powders at the Pt L$_3$ and Ni K-edge simultaneously. $S_0^2$ was fixed at 0.84 and 0.68 for Pt and Ni, respectively as obtained by fitting the reference foils. Fits were done in *R*-space, $k^{1,2,3}$ weighting. 1.2 < *R* < 3.1 Å and Δ*k* = 3.08 − 13.39 Å$^{-1}$ were used for fitting the Pt L$_3$-edge data, and 1.3 < R < 3.1 Å and $\Delta k$ = 2.56 – 11.40 Å$^{-1}$ were used for fitting the Ni K-edge data. The fitting results of the E$_0$ at the Pt L$_3$ and Ni K edges are 8±2 eV and -6±1 eV, respectively.

Calculation of the coordination number, facet orientation and surface concaveness

We used custom MATLAB scripts to measure the CN, facet orientation and surface concaveness for all the atomic sites. We defined the nearest-neighbor distance by fitting the valley of the first and second peak of the pair distribution function for each nanocatalyst. The CN was obtained by counting the number of the nearest-neighbor sites within the cutoff distance. Each atom was classified as a surface site if CN < 12 and as an internal site if CN = 12. To find the facet orientation, we derived a density matrix for each nanocatalyst by convolving the atomic structure with a 3D Gaussian function ($\sigma$ = 4 Å). For each surface site, a normal vector was calculated by computing the gradient of the density matrix at that site. By comparing the normal vectors to the crystallographic directions, we determined the facet orientation of the nanocatalyst. To quantify the surface concaveness, we estimated the surface curvature for all surface sites by using a procedure published elsewhere (*45*).

The normalized local bond orientational order (BOO) parameter

From the 3D atomic model of each nanoparticle, we calculated the local BOO parameters ($Q_4$ and $Q_6$), which are described elsewhere (*31*). The $Q_4$ and $Q_6$ order parameters were computed up to the second shell with a shell radius of 3.5 Å. We then defined the normalized local BOO parameter as $\sqrt{Q_4^2 + Q_6^2}/\sqrt{Q_{4\,fcc}^2 + Q_{6\,fcc}^2}$, where $Q_{4\,fcc}$ and $Q_{6\,fcc}$ are the $Q_4$ and $Q_6$ values of a perfect fcc lattice. The normalized BOO parameter is between 0 and 1, where 0 means $Q_4 = Q_6 = 0$, and 1 represents a perfect fcc crystal structure.

The chemical order parameter

The chemical order of each nanoparticle was calculated by the pair-wise multicomponent short-range order parameter (*40*),

$$\alpha_{ij} = \frac{p_{ij} - C_j}{\delta_{ij} - C_j} \quad (2)$$

where $p_{ij}$ is the probability of finding a $j$-type atom around an $i$-type atom in the first nearest neighbor shell. $C_j$ is the concentration of $j$-type atoms in the nanoparticle and $\delta_{ij}$ is the Kronecker delta function. Since there are primarily Pt and Ni atoms in the nanoparticles, all the $\alpha_{ij}$ parameters are correlated. In this study, we chose $\alpha_{12}$ to represent the chemical order parameter. A positive and a negative $\alpha_{12}$ represent segregation and alloying, respectively.

DFT calculations

The DFT calculations were conducted by the VASP package (*47-50*). The core electrons were described by the projector-augmented-wave method (*51*) and the valence states by plane waves up to 400eV. The exchange-correlation interaction of electrons was defined by the Perdew-Bruke-Ernzerhof functional (*52*). Spin-polarized calculations were used throughout this manuscript for the PtNi nanocatalysts. The Brillouin zone was sampled by a uniform density of 0.19 Å$^{-1}$ in each reciprocal direction. For isolated clusters, only Γ point was considered. A database of OH adsorption energies on the surface Pt sites of the PtNi model catalysts were computed by DFT. The database includes diverse 3D atomic models consisting of nanoclusters and slabs. The nanoclusters of different sizes were built from truncated octahedra, and the slab models were created from closed-packing surfaces, including fcc(110) surfaces, fcc(100) surfaces and concave shapes similar to these published elsewhere (*3*). Ni atoms were introduced by randomly replacing the Pt atoms. The Ni concentration ranges from 0 to 69.6% in the 3D atomic models and from 0 to 83.9% in the local environment (defined as the atoms within the 6.5 Å radius from an adsorption site). Different lattice constants were used to represent tensile and compressive strain. In total, the OH was adsorbed on 207 Pt sites with different Pt CNs, Ni concentration and local environments. The OH binding energy on the Pt sites ($BE_{OH}$) was computed by,

$$BE_{OH} = E_{OH@Model} - E_{OH} - E_{Model} \quad (3)$$

where $E_{OH@Model}$ is the total electronic energy of the optimized OH adsorbed structure, $E_{OH}$ is the electronic energy of the OH radical in the gas phase, and $E_{Model}$ is the energy of optimized model without OH adsorption.

Evaluation of the OH binding energy by DFT-trained ML

To evaluate $BE_{OH}$ for all the surface Pt sites, we used a ML method - the Gaussian process regression (GPR) (*53*). The ML-GPR method was trained by the DFT-calculated $BE_{OH}$, and the local atomic environment of the Pt sites was characterized by the smooth overlap of atomic positions (SOAP) approach (*54, 55*). The cutoff radius of 6.5 Å was selected in SOAP, which was later validated by the accurate prediction of the ML-GPR method (Fig. S11). The GPR was implemented by the Python-scikit package (*56*). The kernel function is defined as normalized polynomial kernel of the partial power spectrum,

$$K(\mathbf{d}_1, \mathbf{d}_2) = \left( \frac{\mathbf{d}_1^T \mathbf{d}_2}{\sqrt{\mathbf{d}_1^T \mathbf{d}_1 \mathbf{d}_2^T \mathbf{d}_2}} \right)^\zeta \quad (4)$$

Where $K(\mathbf{d}_1, \mathbf{d}_2)$ is the kernel function between SOAP descriptors $\mathbf{d}_1$ and $\mathbf{d}_2$, and $\zeta$ is a hyperparameter. In this study, we chose $\zeta = 4$ by balancing the accuracy and transferability. We trained the GPR by randomly choosing 134 atomic models and then used the ML method to predict the $BE_{OH}$ of the remaining 73 atomic models. The RMSE is 0.05 and 0.07 eV per site for the training and test set, respectively (fig S11). The small REMS values indicate the robustness of the ML method. After validating DFT-trained ML, we applied it to evaluate the $BE_{OH}$ of the surface Pt sites of the 7 PtNi AA and 4 Mo-Pt/Ni AA nanocatalysts.

Estimation of the ORR activity based on the OH binding energy

From the ML-identified $BE_{OH}$, we estimated the ORR activity of each surface Pt site by calculating $\Delta BE_{OH}$,

$$\Delta BE_{OH} = BE_{OH} - BE_{OH,Pt(111)} \quad (5)$$

where $BE_{OH,Pt(111)}$ is the OH binding energy of Pt(111). As the ORR activity and $\Delta BE_{OH}$ are related to each other by the volcano-type plot (*32, 33, 57*)), we evaluated the current density of the ORR oxygen using formulas published elsewhere (*32*). On the left side of the volcano plot, we computed the current density of the ORR for the $i^{th}$ surface Pt site ($j_i$) by,

$$kT \ln\left(\frac{j_i}{j_{Pt(111)}}\right) = \Delta BE_{OH,i} \quad (6)$$

where $k$ is the Boltzmann constant, $T$ the temperature, $j_{Pt(111)}$ the current density of the ORR for Pt(111), and $\Delta BE_{OH,i}$ the OH binding energy difference (Eq. (5)) for the $i^{th}$ surface Pt site. We computed the current density of the ORR on the right side of the volcano plot by,

$$kT \ln\left(\frac{j_k}{j_{Pt(111)}}\right) = 0.26 - 0.97 \cdot \Delta BE_{OH,k} \quad (7)$$

where $k$ represents the $k^{th}$ surface Pt site. Based on Eqs. (6) and (7), the current density of any surface Pt site is obtained by,

$$j = \min(j_i, j_k) . \quad (8)$$

Figure 3J shows the ML-identified activity of all the surface Pt sites (blue dots) for the 7 PtNi AA and 4 Mo-Pt/Ni AA nanocatalysts. We observed that the ORR activity of the various surface Pt atoms differs by several orders of magnitude (Fig. 3B, C and Fig. S12). The average activity of these nanocatalysts is in good agreement with the electrochemically measured activity (Fig. 3A and Fig. S2.

Derivation of the local environment descriptor (LED)

We derived LED by fitting a number of experimentally measured structural and chemical factors to the $\Delta BE_{OH}$ of the surface Pt sites of the 7 PtNi AA and 4 Mo-Pt/Ni AA nanocatalysts. We took into account the following factors as the fitting variables: the CN, the surface CN, the sub-surface CN, the average Pt-Pt bond length around each Pt site ($\bar{d}_{Pt}$), the structure / chemical order parameter, the nearest-neighbor Pt and Ni atoms of each surface Pt atom ($NN^{Pt}$ and $NN^{Ni}$), the generalized CN ($\overline{CN}$), the element-based $\overline{CN}$ ($\overline{CN}^{Pt}$ and $\overline{CN}^{Ni}$), etc. We examined the 2, 3 and 4 variable LED equations to minimize the RMSE by,

$$RMSE = \sqrt{\frac{\sum_{i=1}^{N}(\Delta BE_{OH,i}^{ML} - \Delta BE_{OH,i}^{Cal})^2}{N}} \quad (9)$$

where $\Delta BE_{OH,i}^{ML}$ is the $\Delta BE_{OH}$ of the $i^{th}$ surface Pt site obtained by ML, N is the total number of the surface Pt sites, and $\Delta BE_{OH,i}^{Cal} = E_1 * LED - E_0$ with $E_1$ and $E_0$ are two fitting constants. For the 2 variable LED equation, we examined $LED = a_1 x_1 + a_2 x_2$, where $a_1$ and $a_2$ are two fitting constants, $x_1$ and $x_2$ are two fitting variables. We obtained the smallest RMSE value of 0.117 eV by choosing $x_1 = CN$ and $x_2 = e^{-\bar{d}_{Pt}}$. For the 3 variable LED equation, we found that $LED = a_1 x_1 * x_2 + a_2 x_3$ produces the smallest RMSE value of 0.104 eV with $x_1 = NN^{Pt}$, $x_2 = e^{-\bar{d}_{Pt}}$ and $x_3 = \overline{CN}^{Ni}$, which is smaller than that of using the form $LED = a_1 x_1 * x_2 + a_2 x_3$. For the 4 variable LED equation, we examined $LED = a_1 x_1 * x_2 + a_2 x_3 * x_4$ and $LED = a_1 x_1 * x_2 + a_2 x_3 + a_3 x_4$. After testing all the fitting variables, we obtained the smallest RMSE of 0.104 with $LED = a_1 x_1 * x_2 + a_2 x_3$. Based on these analyses, we chose the 3 variable LED equation and revised it to be $LED = NN^{Pt} \cdot e^{-a_1 \cdot \varepsilon} + a_2 \cdot \overline{CN}^{Ni}$, where $\varepsilon = \frac{\bar{d}_{Pt} - d_0}{d_0}$ and $d_0 = 2.75$ Å is the Pt-Pt bond length of the Pt nanoparticles. The revision only changed the fitting constants, but not the RMSE. With this LED equation, we have $a_1 = 1.985$, $a_2 = 1.075$, $E_1 = 0.189$, $E_0 = 0.1703$, and RMSE = 0.104 eV.

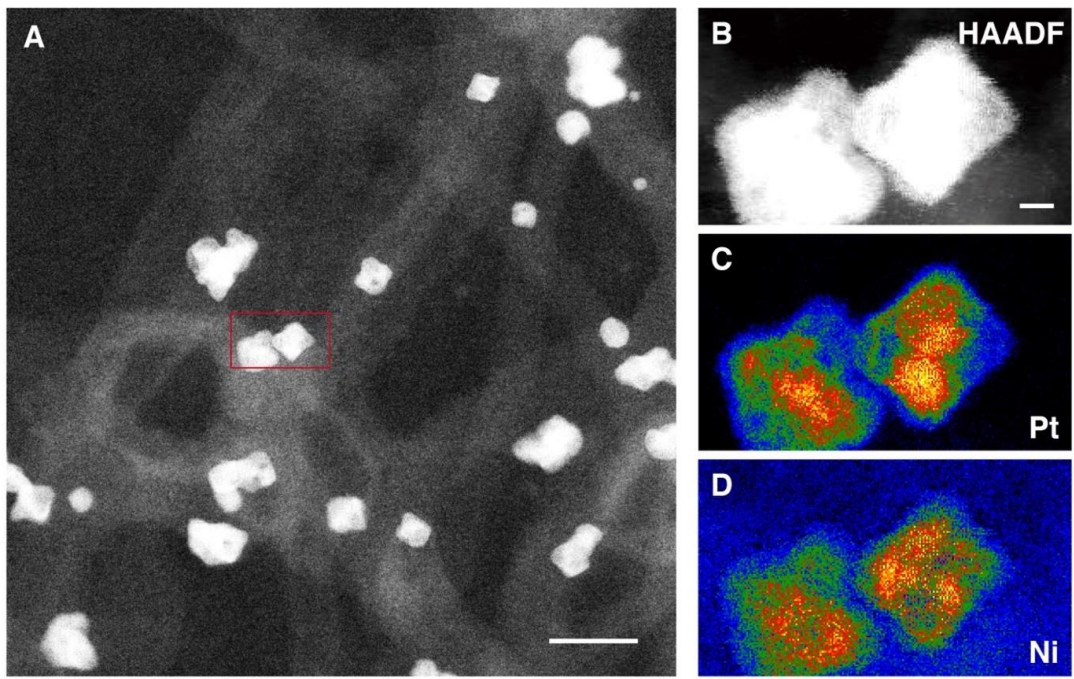

**Fig. S1**. PtNi nanoparticles embedded in a carbon nanotube matrix. (**A**) ADF-STEM image of nanoparticles. (**B**) A magnified image of two PtNi nanoparticles, corresponding to the rectangular region in (**A**). Electron energy loss spectroscopy maps of the distribution of Pt (**C**) and Ni (**D**) in the PtNi nanoparticles, showing that individual Ni but not Pt atoms are distributed around the nanoparticles. Scale bars, 20 nm in (**A**) and 2 nm in (**B**)

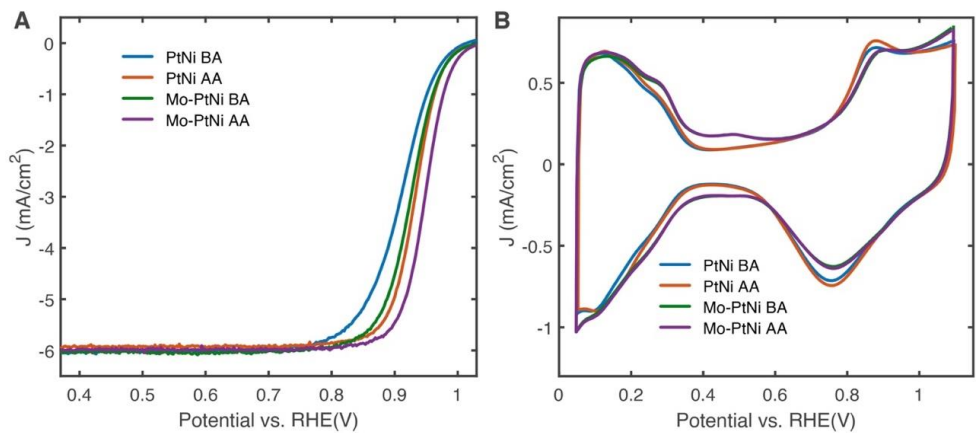

**Fig. S2.** Electrochemical measurements of the nanocatalysts. (**A**) ORR polarization curves of the PtNi BA, PtNi AA, Mo-PtNi BA and Mo-PtNi AA nanoparticles with and without cyclic voltammetry (CV) activation. The ORR plots were recorded at room temperature in an $O_2$-saturated 0.1 M $HClO_4$ aqueous solution with a sweep rate of 20 mV/s and a rotation rate of 1600 rpm. (**B**) CVs of PtNi BA, PtNi AA, Mo-PtNi BA and Mo-PtNi AA nanoparticles. The CV plots were recorded after ORR test at room temperature in a $N_2$ saturated 0.1 M $HClO_4$ electrolyte with sweep rate of 100 mV/s.

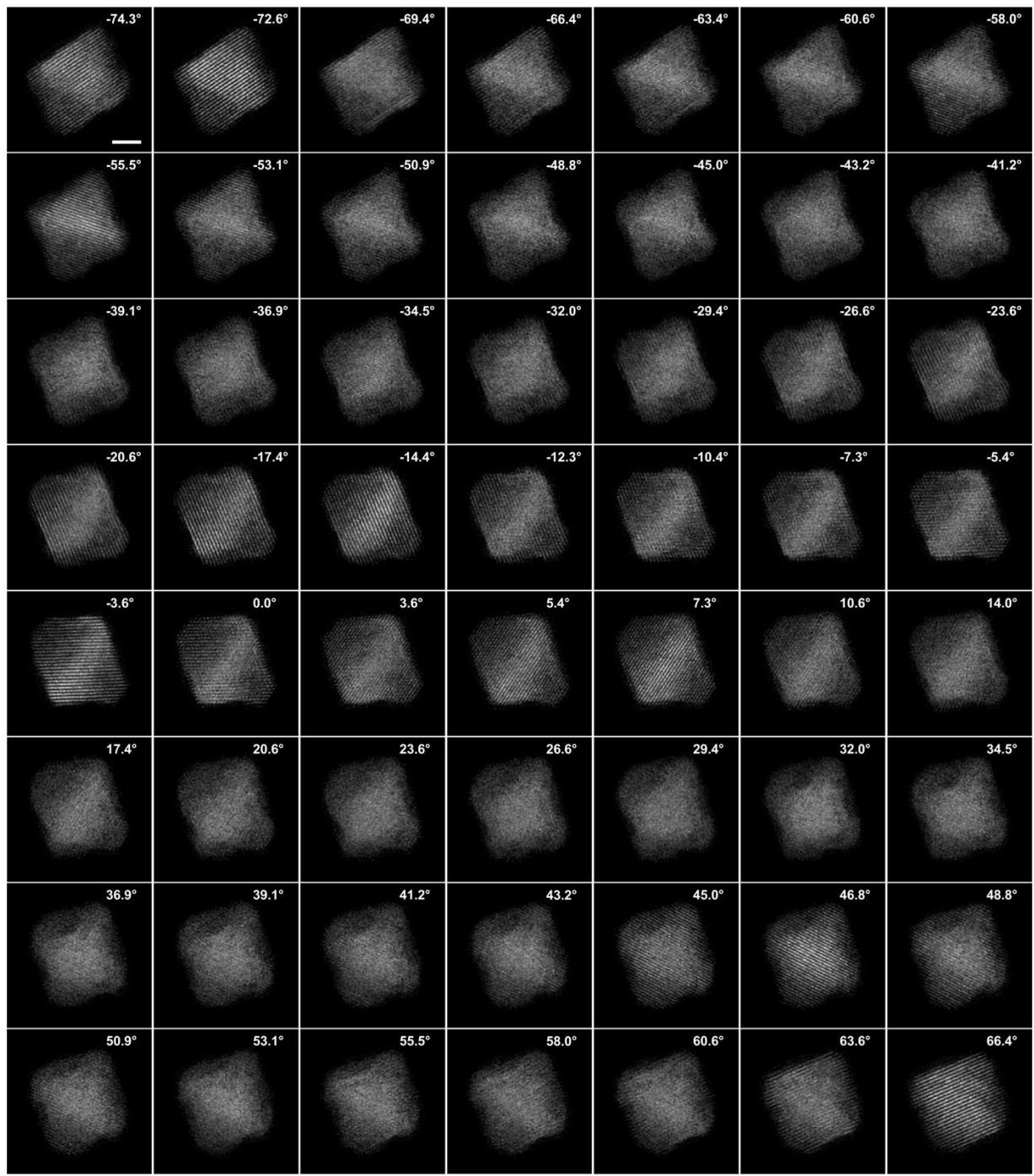

**Fig. S3**. Experimental tomographic tilt series of a PtNi BA nanoparticle (Particle #1), showing 56 ADF-STEM images with a tilt range from -74.3° to +66.4°. The 3D atomic structure and chemical composition of the nanoparticle are shown in Fig. 1 (1st row). Scale bar, 2 nm.

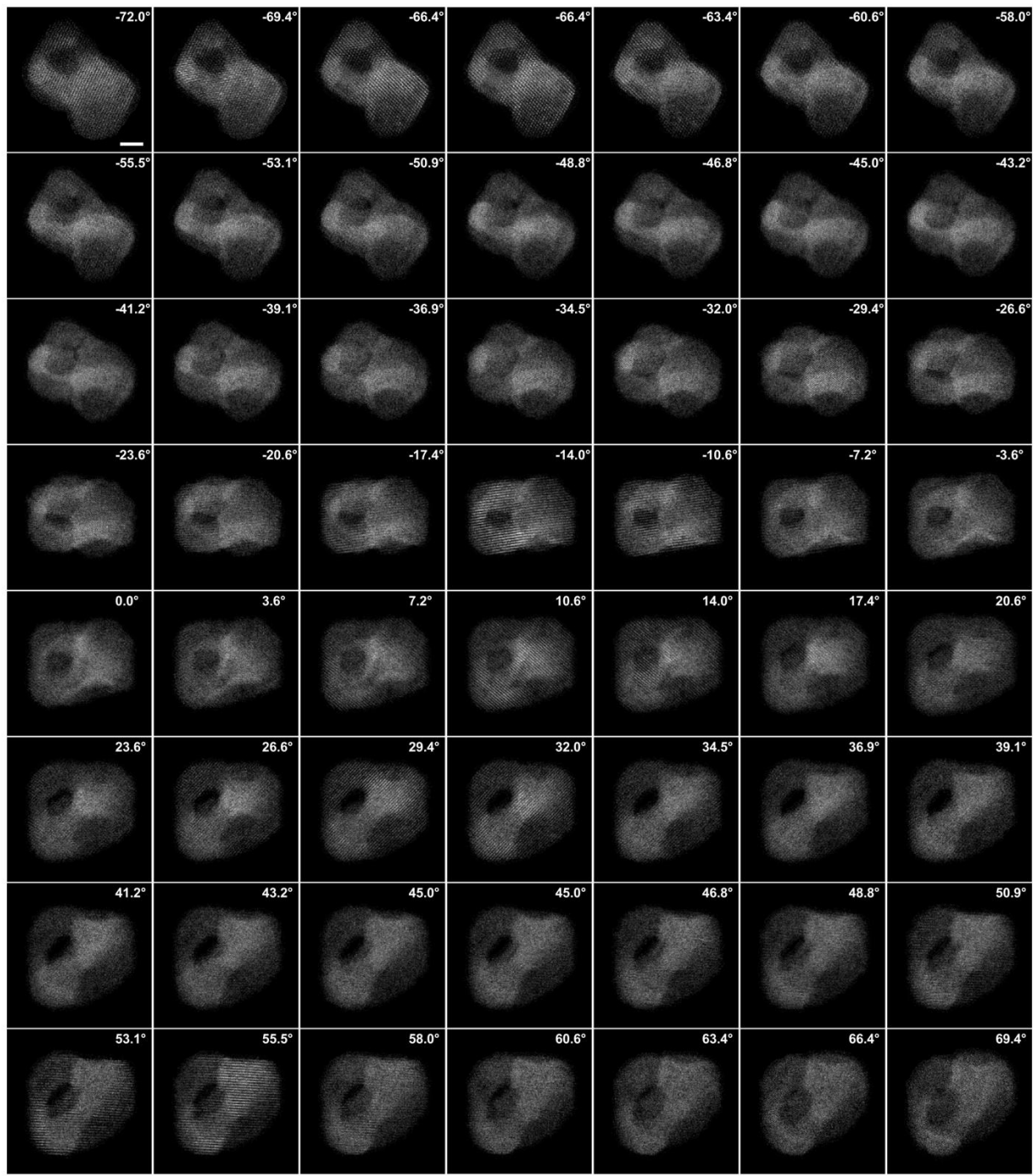

**Fig. S4**. Experimental tomographic tilt series of a PtNi AA nanoparticle (Particle #2), showing 56 ADF-STEM images with a tilt range from -72.0° to +69.4°. The 3D atomic structure and chemical composition of the nanoparticle are shown in Fig. 1 (2nd row). Scale bar, 2 nm.

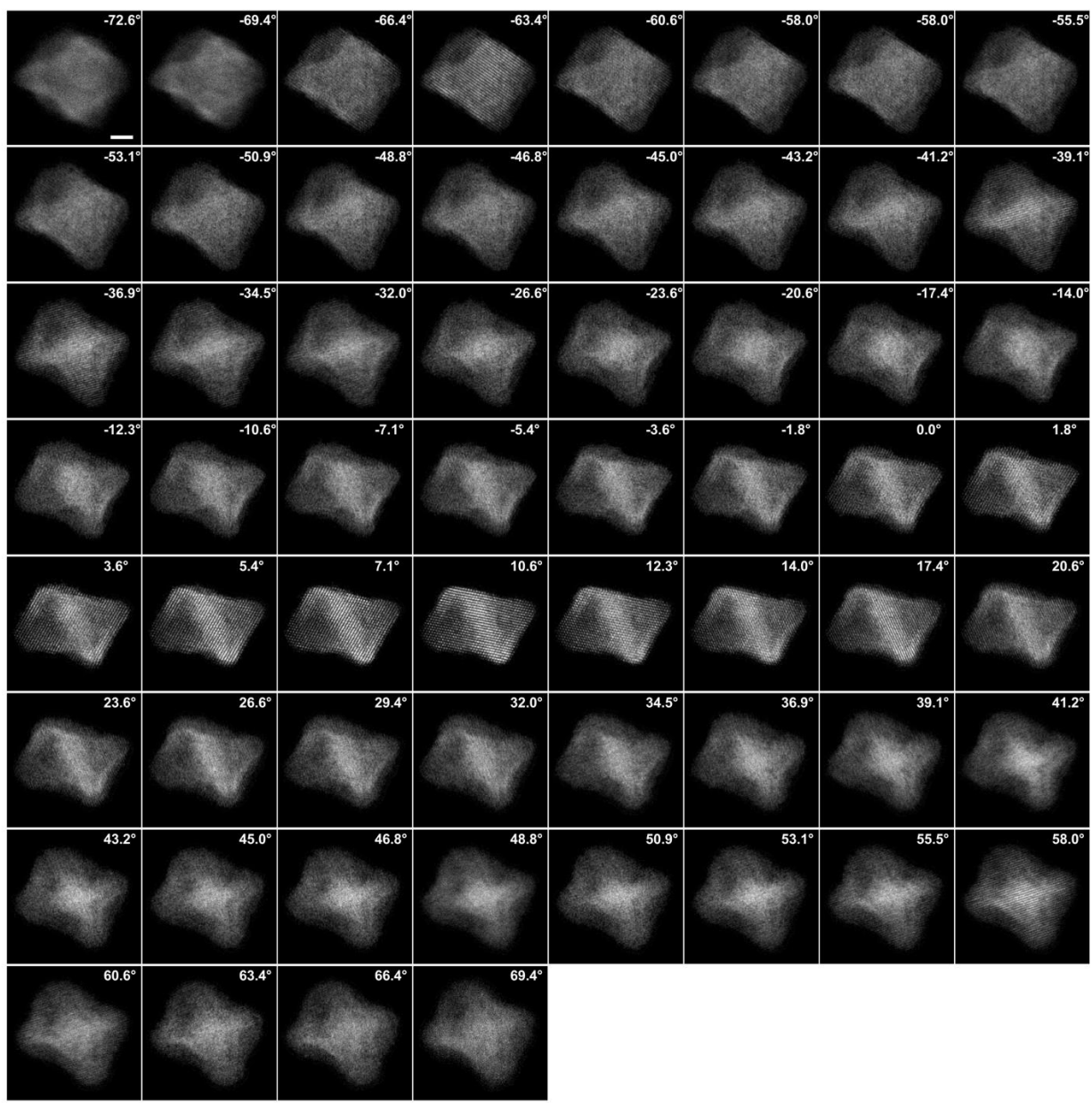

**Fig. S5**. Experimental tomographic tilt series of a Mo-PtNi BA nanoparticle (Particle #3), showing 60 ADF-STEM images with a tilt range from -72.6° to +69.4°. The 3D atomic structure and chemical composition of the nanoparticle are shown in Fig. 1 (3$^{rd}$ row). Scale bar, 2 nm.

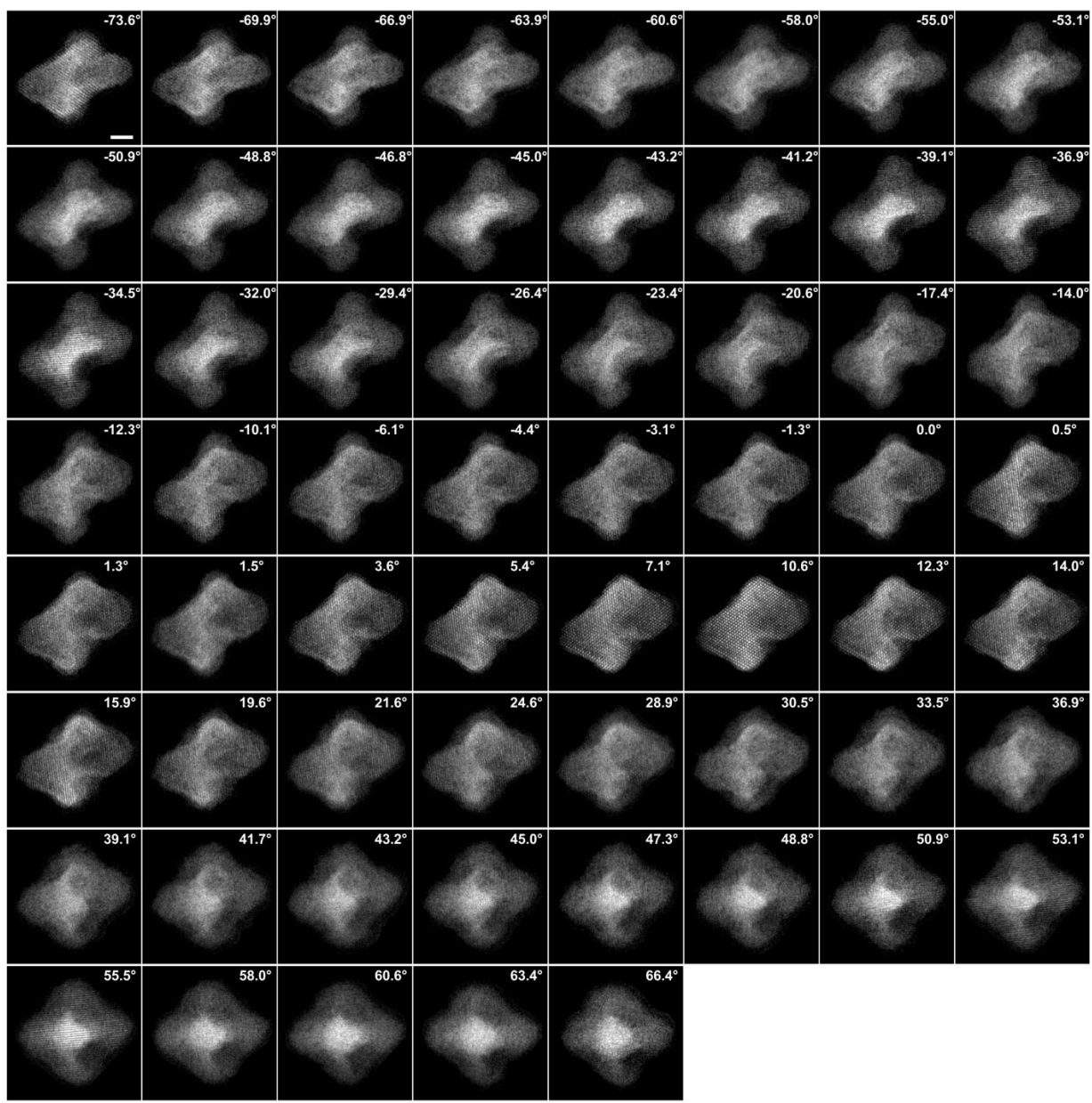

**Fig. S6**. Experimental tomographic tilt series of a Mo-PtNi AA nanoparticle (Particle #4), showing 61 ADF-STEM images with a tilt range from -73.6° to +66.4°. The 3D atomic structure and chemical composition of the nanoparticle are shown in Fig. 1 (4$^{th}$ row). Scale bar, 2 nm.

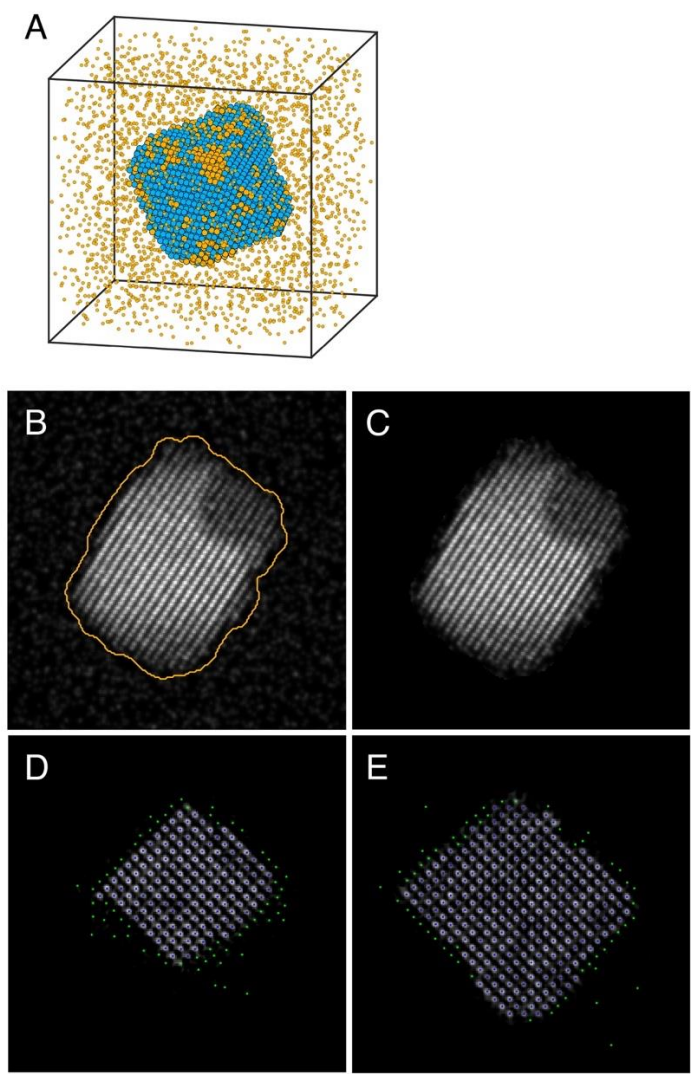

**Fig. S7**. Numerical simulations on the effect of the surrounding Ni atoms on the 3D reconstruction. (**A**) A PtNi atomic model in an environment with individual Ni atoms, where the Ni atom density was estimated from the EELS map. (**B**) A representative projection image calculated from the atomic model, in which the yellow boundary represents a 2D mask determined by the Otsu threshold. (**C**) The projection image after masking. (**D** and **E**) Two representative atomic slices of the reconstructed model, where the blue dots represent atoms consistent with the original model and the green dots show a layer of ghost atoms due to the effect of the Ni environment.

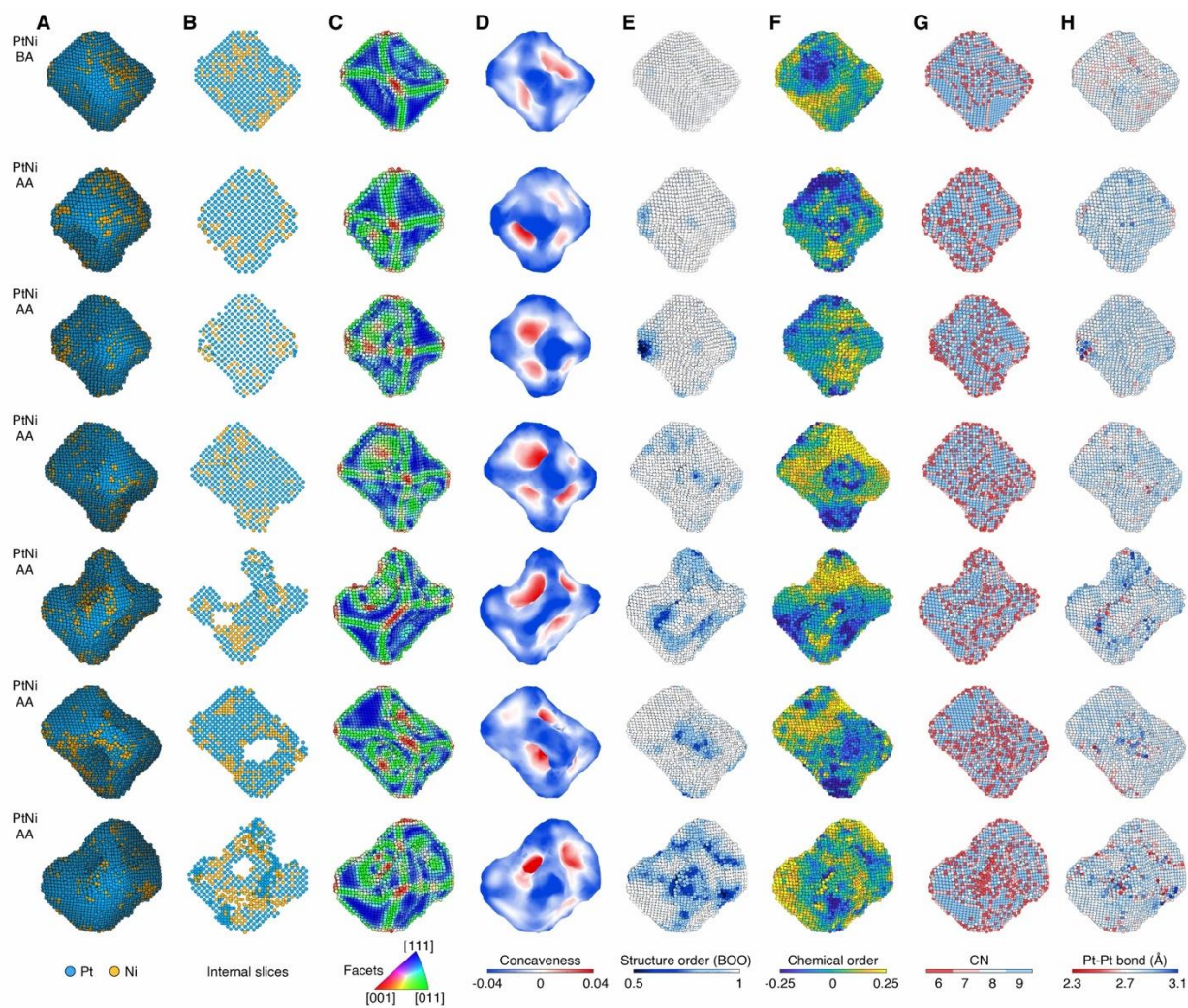

**Fig. S8.** 3D atomic structure and the chemical composition of 7 PtNi nanocatalysts determined by AET. Particles 5-11 correspond to rows 1-7, respectively. From the experimental atomic coordinates, the 3D surface morphology and chemical composition (**A**), elemental segregation in the interior (**B**), facets (**C**), surface concaveness (**D**), structural order/disorder (**E**), chemical order/disorder (**F**), CN (**G**), and surface bonds (**H**) of the nanocatalysts were identified at the single-atom level. For the structural order, BOO = 1 corresponds to a perfect fcc lattice. For the chemical order, positive and a negative numbers represent segregation and alloying, respectively.

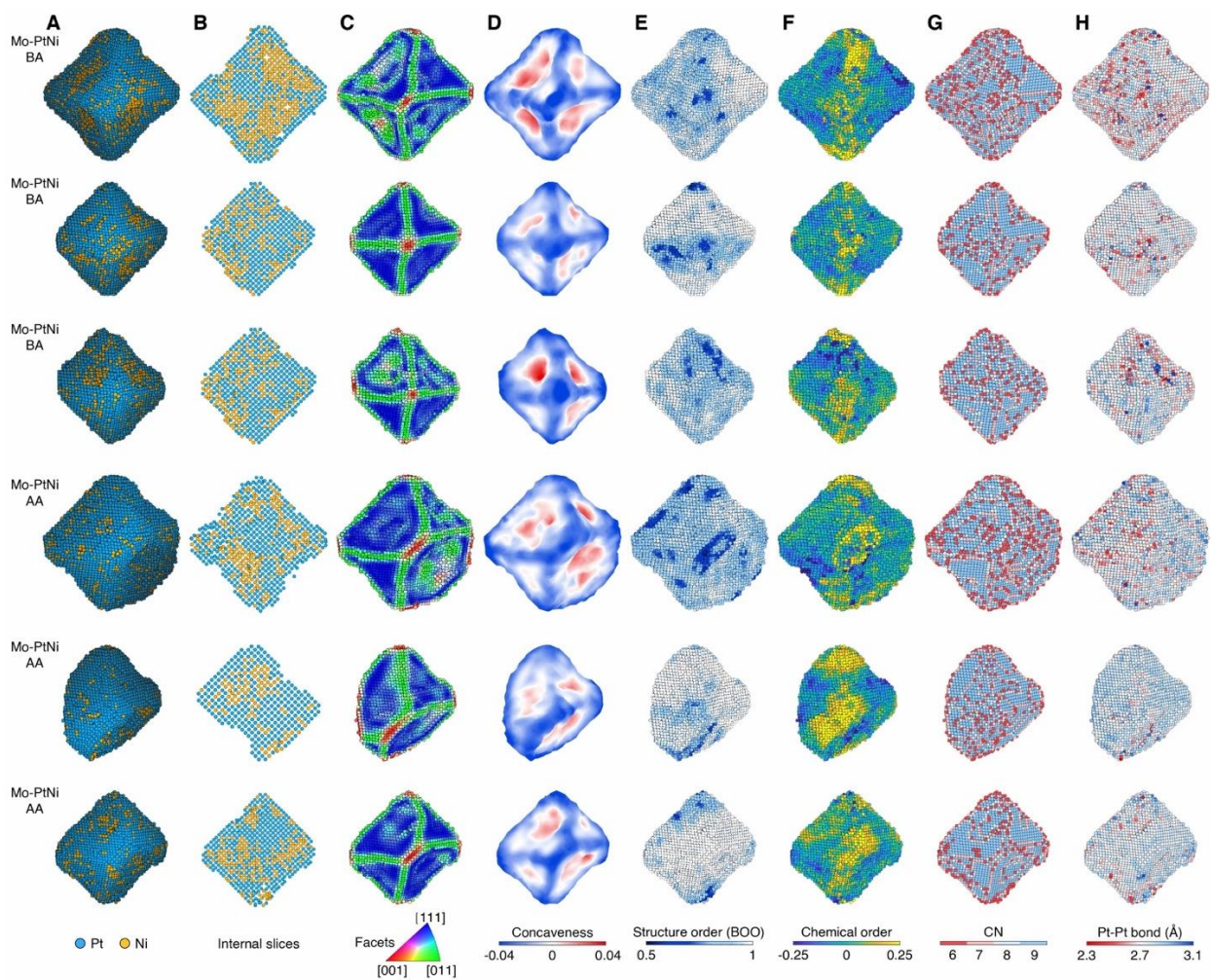

**Fig. S9.** 3D atomic structure and the chemical composition of 6 Mo-PtNi nanocatalysts determined by AET. Particles 12-17 correspond to rows 1-6, respectively. From the experimental atomic coordinates, the 3D surface morphology and chemical composition (**A**), elemental segregation in the interior (**B**), facets (**C**), surface concaveness (**D**), structural order/disorder (**E**), chemical order/disorder (**F**), CN (**G**), and surface bonds (**H**) of the nanocatalysts were identified at the single-atom level.

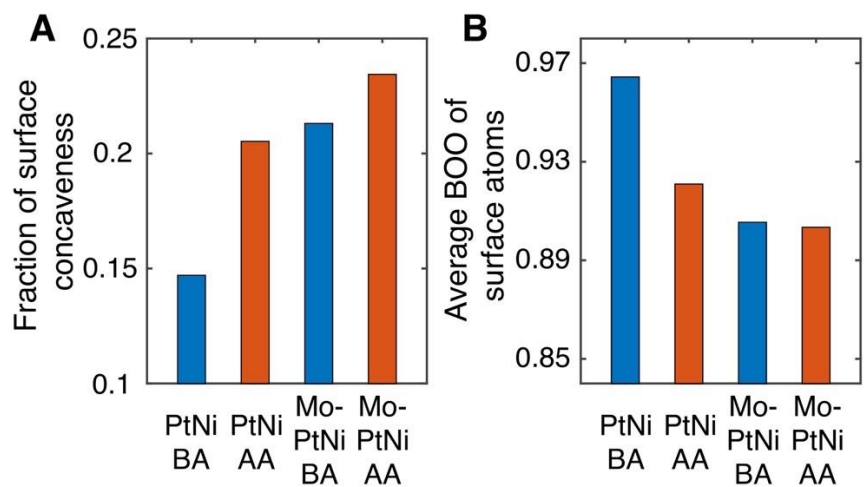

**Fig. S10.** Histogram of the fraction of atoms with surface concaveness (**A**) and the average BOO of the surface atoms (**B**), where BOO = 1 corresponds to a perfect fcc lattice.

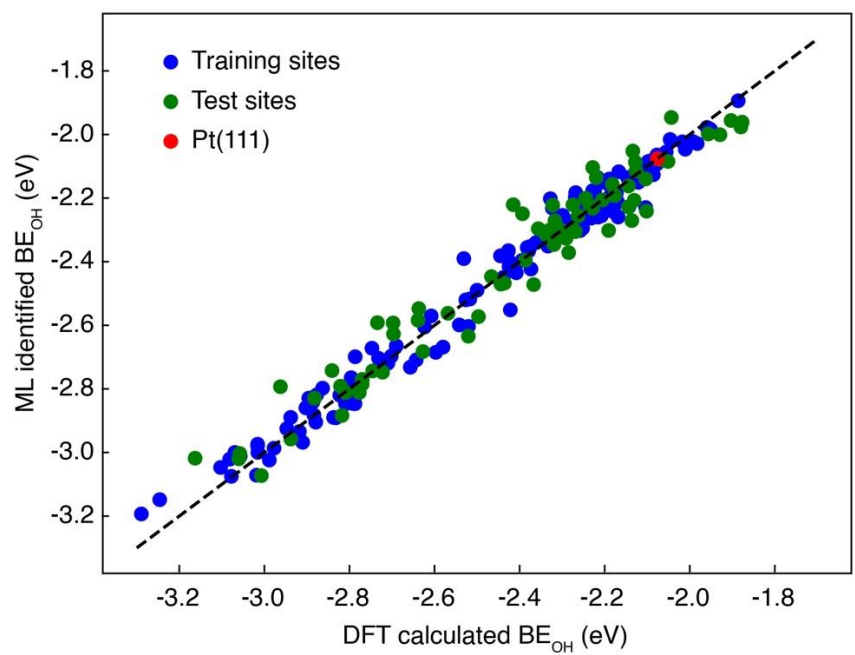

**Fig. S11.** Quantitative comparison between the DFT calculated and ML identified $BE_{OH}$, indicating that ML accurately predicted the $BE_{OH}$ with a root mean square error (RMSE) of 0.05 and 0.07 eV per site for the 134 training and 73 test Pt sites.

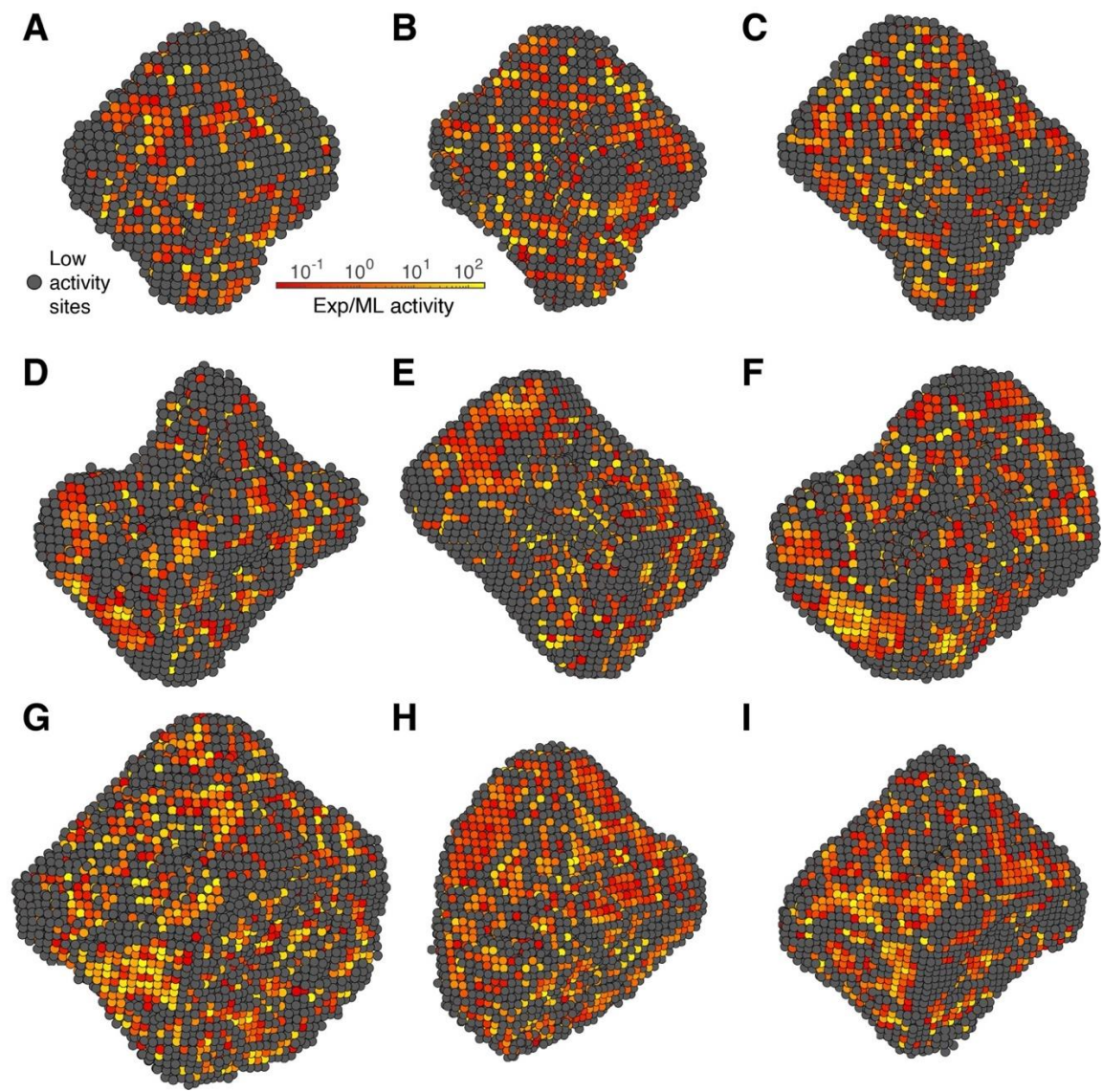

**Fig. S12.** The ORR activity distribution of the surface Pt sites of the PtNi AA and Mo-PtNi AA nanocatalysts, derived by ML using the experimentally measured 3D atomic coordinates as input. (**A-F**) Particles 6-11 of PtNi AA, respectively. (**G-I**) Particles 15-17 of Mo-PtNi AA, respectively.

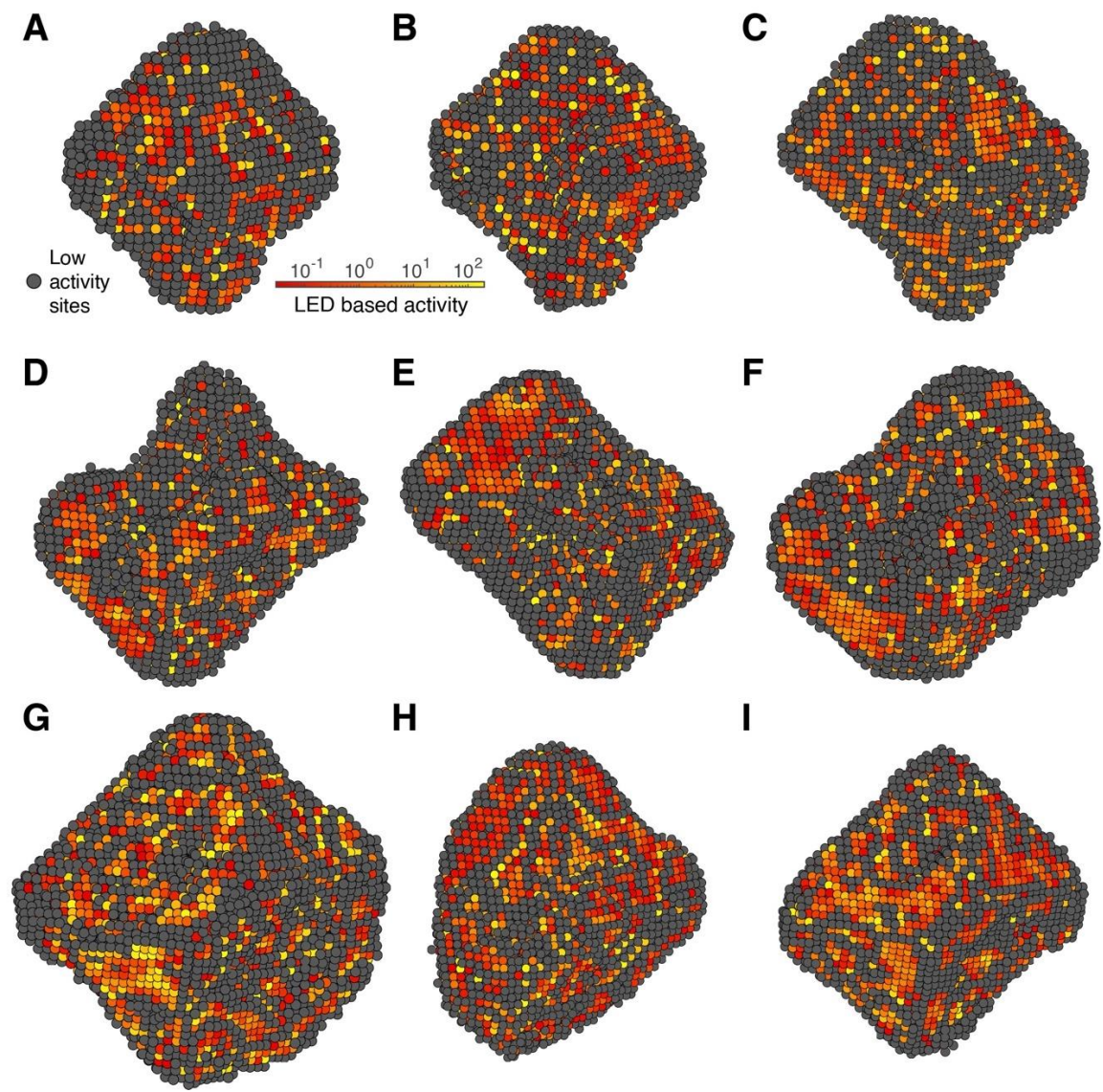

**Fig. S13.** The ORR activity distribution of the surface Pt sites of the PtNi AA and Mo-PtNi AA nanocatalysts based on LED. (**A-F**) Particles 6-11 of PtNi AA, respectively. (**G-I**) Particles 15-17 of Mo-PtNi AA, respectively. The LED-based activity distribution is consistent with that identified by ML (Fig. S12.

**Table S1**. AET data collection, processing, reconstruction, refinement and statistics of PtNi and Mo-PtNi.

| | Particle[a] 1 | Particle 2 | Particle 3 | Particle 4 | Particle 5 | Particle 6 | Particle 7 | Particle 8 | Particle 9 |
|---|---|---|---|---|---|---|---|---|---|
| **Data collection and processing** | | | | | | | | | |
| Voltage (kV) | 300 | 300 | 300 | 300 | 300 | 300 | 300 | 300 | 300 |
| Convergence semi-angle (mrad) | 17.1 | 17.1 | 17.1 | 17.1 | 17.1 | 17.1 | 17.1 | 17.1 | 17.1 |
| Probe size (Å) | 0.7 | 0.7 | 0.7 | 0.7 | 0.7 | 0.7 | 0.7 | 0.7 | 0.7 |
| Detector inner angle (mrad) | 30 | 30 | 30 | 30 | 30 | 30 | 30 | 30 | 30 |
| Detector outer angle (mrad) | 195 | 195 | 195 | 195 | 195 | 195 | 195 | 195 | 195 |
| Depth of focus (nm) | 14 | 14 | 14 | 14 | 14 | 14 | 14 | 14 | 14 |
| Pixel size (Å) | 0.467 | 0.467 | 0.467 | 0.467 | 0.467 | 0.467 | 0.467 | 0.467 | 0.467 |
| # of images | 56 | 56 | 60 | 61 | 56 | 57 | 55 | 53 | 56 |
| Tilt range (°) | -74.3° 66.4° | -72.0° 69.4° | -72.6° 69.4° | -73.6° 66.4° | -72.6° 66.4° | -74.3° 66.4° | -72.6° 69.4° | -72.6° 69.4° | -74.3° 63.4° |
| Total electron dose ($10^5$ e/Å$^2$) | 7.8 | 7.8 | 8.4 | 8.5 | 7.8 | 7.9 | 7.7 | 7.4 | 7.8 |
| **Reconstruction** | | | | | | | | | |
| Algorithm | RESIRE | RESIRE | RESIRE | RESIRE | RESIRE | RESIRE | RESIRE | RESIRE | RESIRE |
| Oversampling ratio | 4 | 4 | 4 | 4 | 4 | 4 | 4 | 4 | 4 |
| Number of iterations | 200 | 200 | 200 | 200 | 200 | 200 | 200 | 200 | 200 |
| **Refinement** | | | | | | | | | |
| $R_1$ (%)[b] | 19.41 | 21.84 | 14.62 | 16.74 | 19.29 | 22.41 | 19.81 | 18.33 | 22.33 |
| R (%)[c] | 8.00 | 7.50 | 6.39 | 8.16 | 8.86 | 7.31 | 6.77 | 6.39 | 7.55 |
| B' factors (Å$^2$) | | | | | | | | | |
| Ni atoms | 10.76 | 27.48 | 10.73 | 16.44 | 6.38 | 10.57 | 11.34 | 7.96 | 10.75 |
| Pt atoms | 20.59 | 35.09 | 25.70 | 42.21 | 14.24 | 15.31 | 26.95 | 25.53 | 23.94 |
| **Statistics** | | | | | | | | | |
| # of atoms | | | | | | | | | |
| Total | 7235 | 11561 | 13618 | 10937 | 6704 | 4281 | 4383 | 6435 | 6927 |
| Ni | 2184 | 3239 | 5241 | 3588 | 1877 | 850 | 854 | 1442 | 1795 |
| Pt | 5051 | 8322 | 8377 | 7349 | 4827 | 3431 | 3529 | 4993 | 5132 |
| Pt/Ni | 2.31 | 2.57 | 1.60 | 2.05 | 2.57 | 4.04 | 4.13 | 3.46 | 2.86 |

[a] Particles 1-4 correspond to those shown in rows 1-4 in Fig. 1, respectively. Particles 5-11 correspond to rows 1-7 in Fig. S8, respectively. Particles 12-17 correspond to rows 1-6 in Fig. S9, respectively.
[b] The $R_1$ factor is defined as equation 5 in ref. (*23*). Same in Table S2.

[c] The R factor is defined in equation 4 in ref. (*31*). Same in Table S2.

**Table S2**. AET data collection, processing, reconstruction, refinement and statistics of PtNi and Mo-PtNi

|  | Particle 10 | Particle 11 | Particle 12 | Particle 13 | Particle 14 | Particle 15 | Particle 16 | Particle 17 |
|---|---|---|---|---|---|---|---|---|
| **Data collection and processing** | | | | | | | | |
| Voltage (kV) | 300 | 300 | 300 | 300 | 300 | 300 | 300 | 300 |
| Convergence semi-angle (mrad) | 17.1 | 17.1 | 17.1 | 17.1 | 17.1 | 17.1 | 17.1 | 17.1 |
| Probe size (Å) | 0.7 | 0.7 | 0.7 | 0.7 | 0.7 | 0.7 | 0.7 | 0.7 |
| Detector inner angle (mrad) | 30 | 30 | 30 | 30 | 30 | 30 | 30 | 30 |
| Detector outer angle (mrad) | 195 | 195 | 195 | 195 | 195 | 195 | 195 | 195 |
| Depth of focus (nm) | 14 | 14 | 14 | 14 | 14 | 14 | 14 | 14 |
| Pixel size (Å) | 0.467 | 0.467 | 0.467 | 0.467 | 0.467 | 0.467 | 0.467 | 0.467 |
| # of images | 54 | 54 | 58 | 60 | 59 | 59 | 57 | 57 |
| Tilt range (°) | -74.3° 66.4° | -74.3° 66.4° | -72.6° 69.4° | -72.6° 69.4° | -71.0° 69.4° | -72.6° 66.4° | -72.6° 66.4° | -72.6° 66.4° |
| Total electron dose ($10^5$ e/Å$^2$) | 7.5 | 7.5 | 8.1 | 8.4 | 8.2 | 8.2 | 7.9 | 7.9 |
| **Reconstruction** | | | | | | | | |
| Algorithm | RESIRE | RESIRE | RESIRE | RESIRE | RESIRE | RESIRE | RESIRE | RESIRE |
| Oversampling ratio | 4 | 4 | 4 | 4 | 4 | 4 | 4 | 4 |
| Number of iterations | 200 | 200 | 200 | 200 | 200 | 200 | 200 | 200 |
| **Refinement** | | | | | | | | |
| $R_1$ (%) | 19.61 | 19.83 | 14.75 | 15.41 | 16.00 | 16.42 | 18.39 | 15.68 |
| R (%) | 8.07 | 7.73 | 6.90 | 6.26 | 6.50 | 6.03 | 6.12 | 6.34 |
| B' factors (Å$^2$) | | | | | | | | |
| Type 1 atoms | 12.97 | 11.82 | 7.36 | 9.55 | 21.29 | 10.11 | 7.79 | 7.41 |
| Type 2 atoms | 20.75 | 29.90 | 22.93 | 28.24 | 42.00 | 39.18 | 28.70 | 26.83 |
| **Statistics** | | | | | | | | |
| # of atoms | | | | | | | | |
| Total | 8082 | 9275 | 15694 | 9532 | 7220 | 14172 | 8619 | 10589 |
| Ni | 2005 | 3190 | 6192 | 3481 | 2166 | 5696 | 2594 | 3937 |
| Pt | 6077 | 6085 | 9502 | 6051 | 5054 | 8476 | 6025 | 6652 |
| Pt/Ni | 3.03 | 1.91 | 1.53 | 1.74 | 2.33 | 1.49 | 2.32 | 1.69 |

**Table S3**. Electrochemical measurements of the specific activity, mass activity, and electrochemically active surface area (ECSA) of the four types of the nanocatalysts.

| Nanocatalyst | Specific Activity (mA/cm$^2$) | Mass Activity (mA/µg$_{PGM}$) | ECSA (m$^2$/g$_{PGM}$) |
|---|---|---|---|
| PtNi BA | 2.9 | 1.1 | 38 |
| PtNi AA | 4.8 | 1.9 | 39 |
| Mo-PtNi BA | 3.3 | 1.2 | 36 |
| Mo-PtNi AA | 9.3 | 3.5 | 38 |

**Table S4**. The Pt-Pt bond lengths of the four nanocatalysts obtained by EXAFS fitting.

|  | PtNi BA | PtNi AA | Mo-PtNi BA | Mo-PtNi AA |
|---|---|---|---|---|
| Pt-Pt bond length (Å) | 2.708 ± 0.006 | 2.735 ± 0.004 | 2.702 ± 0.005 | 2.724 ± 0.002 |